\documentclass[twocolumn,showpacs,groupedaddress,longbibliography]{revtex4-1}

\usepackage{xfrac}
\usepackage{subfigure}
\usepackage{bm}
\usepackage{graphicx}
\usepackage{amsfonts}
\usepackage{amssymb}
\usepackage{amsmath}
\usepackage{xcolor}
\usepackage{color}      
\usepackage{soul} 

\definecolor{acolor}{rgb}{1,0.6275,0}
\definecolor{bcolor}{rgb}{0.4392,0.1882,0.6275}
\definecolor{column}{rgb}{0.0941,0.6039,0.9922}
\definecolor{line}{rgb}{0.8314,0,0.1137}
\definecolor{bluematlab}{rgb}{0,0.4470,0.7410}

\begin{document}

\title{Geometry and elasticity of a knitted fabric}

\author{Samuel Poincloux\textsuperscript{1}}
\author{Mokhtar Adda-Bedia\textsuperscript{2}} 
\author{Fr\'ed\'eric Lechenault\textsuperscript{1}}
\affiliation{\textsuperscript{1}Laboratoire de Physique Statistique, Ecole Normale Sup\'erieure, PSL Research University, CNRS, F-75231 Paris, France}
\affiliation{\textsuperscript{2}Universit\'e de Lyon, Ecole Normale Sup\'erieure de Lyon, Universit\'e Claude Bernard, CNRS, Laboratoire de Physique, F-69342 Lyon, France}

\begin{abstract}
Knitting is not only a mere art and craft hobby but also a thousand year old technology. Unlike weaving, it can produce loose yet extremely stretchable fabrics with almost vanishing rigidity, a desirable property exhibited by hardly any bulk material. It also enables the engineering of arbitrarily shaped two and three-dimensional objects with tunable mechanical response. In contrast with the extensive body of related empirical knowledge and despite a growing industrial interest, the physical ingredients underlying these intriguing mechanical properties remain poorly understood. To make some progress in this direction, we study a model tricot made of a single elastic thread knitted into the common pattern called \textit{stockinette}. On the one hand, we experimentally investigate its tensile response and measure local  displacements of the stitches during deformation. On the other hand, we derive a first-principle mechanical model for the displacement field based on the yarn bending energy, the conservation of its total length and the topological constraints on the constitutive stitches. Our model solves both the shape and mechanical response of the knit and agrees quantitatively with our measurements. This study thus provides a fundamental framework for the understanding of knitted fabrics, paving the way to thread-based smart materials.
\end{abstract}

\maketitle


Due to the wide range of applications, the advancement in knitting technology as well as the availability of high performance fibers, knitted materials are commonly employed in various innovative areas. For instance, they are intensively utilized in textile industry~\cite{Hu2012review}, advanced engineering~\cite{leong2000potential}, biomedical and biomimetic applications~\cite{Gladman2016biomimetic,Haines2016new}. A basic knit consists in a single yarn which is topologically constrained to form intertwined loops, or \textit{stitches}, from which originates its effective dimensionality. While the topological properties of a knitted fabric are usually unaltered, the stitches can undergo large deformations due to their curved nature and the fact that the yarn can slide from one stitch into the neighbouring ones. Those properties manifest also in the outstanding drapability of the resultant knitted fabrics allowing for the shaping of complex curved composite components. Moreover, while the constituent yarn shows significant resistance to elongation, a knitted fabric can endure large strains in response to small applied tractions. Pulling on a typical scarf can easily produce deformations of the order of 100\% while the same force applied on the yarn itself would only deform it by a few percent. A stretched knit also exhibits a characteristic catenary shape similarly to incompressible bulk materials.

Although most of the efforts have been focusing on woven fabrics, the peculiar properties of knitted materials have induced increasing interest in modelling their mechanical behavior. Several early studies have addressed the fact that a knit is comprised of a discrete network of repetitive stitches characterized by a given topology. On the one hand, geometrical models have focused on the geometry of the loops formed by a stitch and the resulting dimensional properties of the fabric~\cite{peirce1947geometrical,hurd1953fundamental,leaf1955,munden1959geometry,popper1966theoretical}. They consist in deriving a set of parameters and equations for modelling the crossing of yarns in a stitch as a set of inextensible curves. On the other hand, mechanical models that take into account both the elasticity of the yarn and the topology of the stitch have been proposed. To assess the equilibrium shape of a stitch and its mechanical properties, many variations of Euler-Bernoulli beam theory or beam and truss models have been proposed over the past years~\cite{leaf1960,shanahan1970theoretical,hepworth1976mechanics,hong2002theoretical,dusserre2014elastic,de1977energy,choi2003energy,choi2006shape}. Although those studies allow for modelling local equilibrium configurations and mechanical properties of a single stitch, they do not describe the mechanics of a whole fabric unless  homogeneous deformations are assumed. Recently, new approaches emerged that describe the fabric as a grid where stitches are base units~\cite{wu1994computer,loginov2002modellingI,yuksel2012stitch}. Each stitch is sub-divided into constituent elements; each element represents a mechanical equivalent of yarn when it undergoes deformations and length redistribution during the fabric extension. The deformation of the whole fabric is determined by imposing boundary conditions and kinematic relationships between adjacent cells. Finally, robust and efficient cloth simulation has long been a research focus in the computer graphics community. Progress in numerical modeling allowed for the development of yarn-based simulation techniques for realistic and efficient dynamic simulation of knitted clothing and their mechanical modelling~\cite{chen2003realistic,kaldor2008simulating}. However, producing the required yarn-level models remains a challenge for this type of approach.

Despite a long history of domestic use and intensive industrial applications, few approaches have focused on deriving the mechanical properties of knits and their morphology from fundamental principles. To this purpose, we crafted a fabric using a model elastic yarn knitted into a stockinette stitch pattern~\cite{Anbumani2007knitting}. Then, we implemented tensile test experiments under different loading configurations to measure the mechanical response of the fabric, while monitoring its shape using high precision imaging. Furthermore, we developed a two dimensional model relying on a description of the stitch field from which we derived the mechanical properties of the whole fabric. Our approach is based on the regular structure of the stitches that imposes topological constraints during the deformation of the fabric. The model invokes scale separation between the yarn diameter and the stitch extension and neglects the stretching and twisting of the yarn. These assumptions allow us to focus on the bending energy of the yarn which, combined with conservation of its total length and kinematic conditions on neighbouring stitches, yields general equations for the mechanics of the fabric. The equations of the model are solved for the corresponding experimental situations and the predicted deformation field is found to be in agreement with experimental data. Although our model is specifically applied to the stockinette stitch pattern, it provides a general framework for the study of a large class of knits.

\section{Experimental study} 
\begin{figure}[tbh]
\centerline{\includegraphics[width=0.9\linewidth]{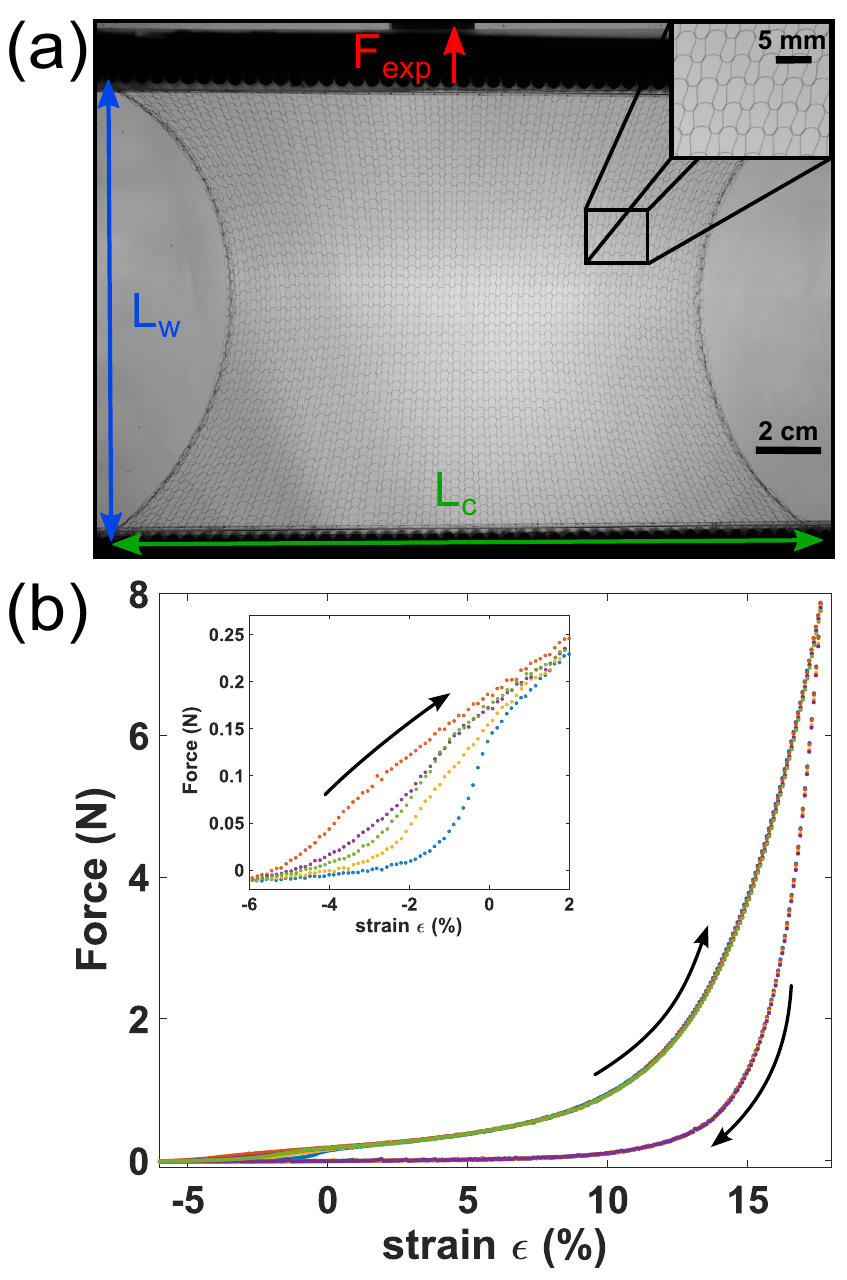}}
\caption{A knitted fabric is stretched along the wale direction. Clamps hold the upper and lower rows preventing any displacement of the corresponding stitches. The mechanical response is probed using a traction bench equipped with a dynamometer and the stitches pattern is imaged through a digital camera (see Materials and Methods and Movie S1 in Supplemental Material). (a) Picture of the deformed stockinette-knit fabric showing the topology of the stitches and their layout in the typical catenary shape. The global dimension of the fabric $L_c$ and $L_w$ as well as the direction of the pulling force are shown. (b) Mechanical response of the fabric over $5$ loading--unloading cycles, each cycle is labelled by a different color. The strain is defined by $\varepsilon=(L_w-L^0_w)/L^0_w$ such that the origin $\varepsilon=0$ corresponds to the extension above which the force signal is reproducible over the cycles. The inset zooms in the force curves close to $\varepsilon=0$ and for different cycles during the loading phase.
\label{fig:experiments}}
\end{figure}

As a model experiment, we use a thin nylon thread and a mechanical knitting machine to manufacture a $51\times51$ stitch fabric in the topologically simplest knit pattern known as stockinette, or "point Jersey" (see Fig.~\ref{fig:experiments}(a)). In the plane of the fabric, stitches are organised along rows and columns and the corresponding directions are usually called \textit{course} and \textit{wale} respectively. During mechanical tests, the fabric is stretched along the wale direction while clamped along the course direction using two parallel rows of $51$ equidistant nails with the same spacing as the one between the needles of the knitting machine. In this configuration, the upper and lower rows have a fixed length $L_c^0=227\textrm{mm}$ and the size $L_w$ in the wale direction is varied during the experiment (see Materials and Methods). A sample is cyclically stretched uniaxially up to a maximum extension and then released back to its initial state. As shown in Fig.~\ref{fig:experiments}(b), the mechanical response of the fabric can be separated into two regions, a first one with large variability over the cycles and almost vanishing stiffness, and a second one, starting at $L_w\equiv L^0_w=125\textrm{mm}$, showing a stiffening behavior together with a large hysteresis between loading and unloading phases.

Fig.~\ref{fig:experiments}(b) also shows that the work performed to return to the initial state is nearly half that needed to stretch the fabric, yet the response is consistently elastic and repeatable over the cycles, as the fabric always retrieves its initial shape. This dissipative behavior results from self-friction of the yarn which contributes oppositely in the loading and unloading phases~\cite{dusserre2015modelling}: when one stretches the fabric, the frictional part of the force points down, thus adding to the elastic load, but when one unloads the sample, this force points upwards, thus subtracting from the elastic part of the load. This effect yields different stiffening behavior upon loading and unloading of the fabric. Finally, as far as plastic deformation goes, we have checked that during deformation of the knit, the thread does not undergo irreversible deformation: it remains straight upon un-knitting the fabric after the deformation cycles.

\begin{figure}[tbh]
\centerline{\includegraphics[width=0.9\linewidth]{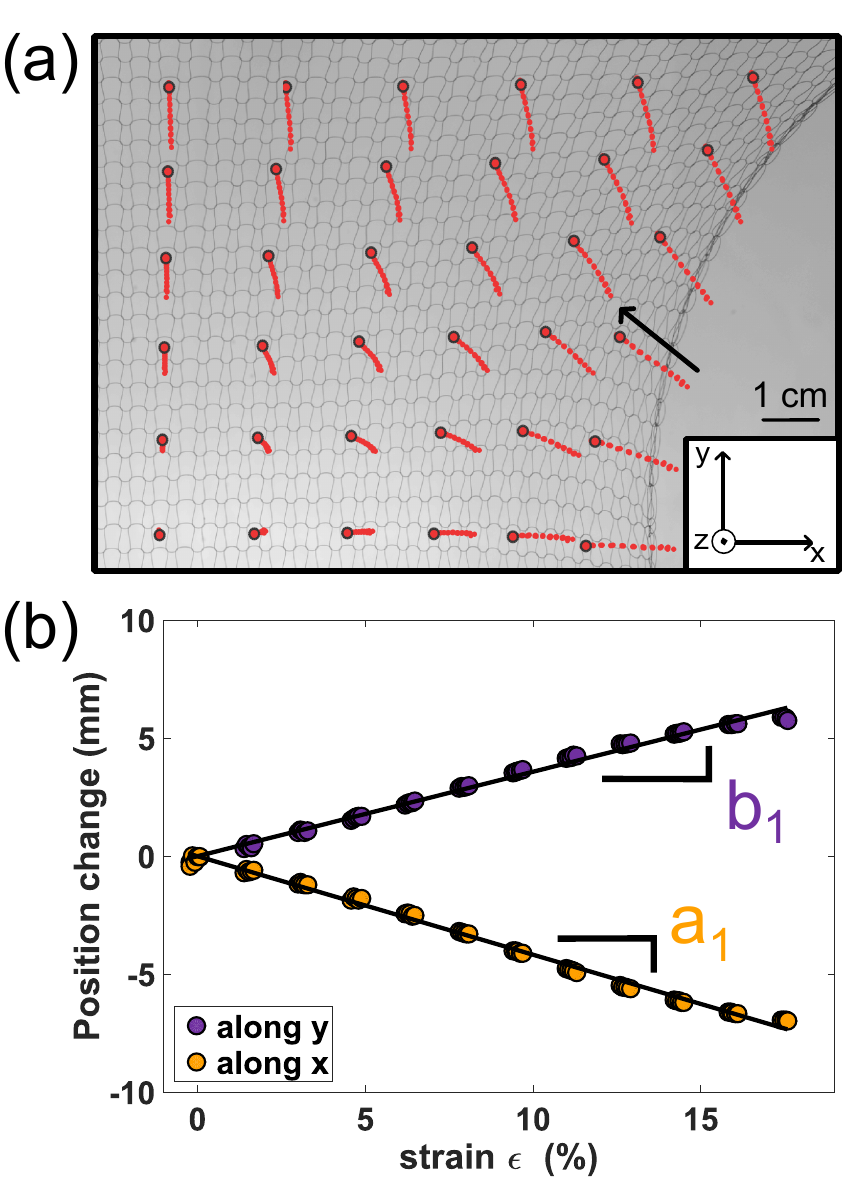}}
\caption{The trajectories of the stitches are tracked while the fabric is stretched. (a) Picture of the upper right part of the fabric for $\varepsilon=18\%$, referred as the actual configuration. Red dots represent the paths of a selection of stitch centroids when the strain $\varepsilon$ is varied from $0$ to $18\%$ during the stretching phase of the $5$ cycles (Movie S2 in Supplemental Material shows the corresponding trajectories). Black circles show the corresponding positions in the actual configuration. (b) Plot of the projections of the change in position of the stitch's centroid next to the black arrow along $x$ (wale) and $y$ (course) direction as function of the applied strain during the stretching phases. Black solid lines represent the best linear fit whose slopes are measures of $a_1$ and $b_1$. \label{fig:stitch_track}}
\end{figure}

Concomitantly to force measurements, the morphology of the fabric is recorded with a high resolution camera. After image segmentation, the geometric center of each stitch is tracked individually during the stretching and unloading phases of the cycles. As shown in Fig.~\ref{fig:stitch_track}(a), a striking feature during the deformation of the fabric is that all the stitches centroids follow quite straight trajectories as long as $\varepsilon\geq0$, where $\varepsilon=(L_w- L^0_w)/L_w^0$ is the global stretching of the fabric. This allows us to write the individual positions of the stitches as $\vec{u}(j,i)=\vec{u}_{0}(j,i)+\varepsilon\vec{u}_{1}(j,i)$ where the indices $(j,i)$ designate the stitch position along the course and wale direction respectively and the vector field $\vec{u}$ is defined with respect to a common reference point in the $(x,y)$-plane of the fabric. For $\varepsilon>0$ and within the whole range of applied stretching, the displacement field of the stitches is described by a strain-independent vector field $\vec{u}_1$ whose components $(a_1,b_1)$ can be retrieved experimentally (see Fig.~\ref{fig:stitch_track}(b)). The same affine behavior of the displacement field is observed during the unloading phase (see Movie S2 in Supplemental Material). This confirms that the hysteretic behavior in the elastic response of the fabric originates from inter-yarn friction and could therefore be included in a global stiffness constant while maintaining the same local displacement field of the stitches in loading and unloading phases.

The vector field $\vec{u}_0$ describes a reference state of the system for which $L_w=L^0_w$. In this configuration, the fabric is already deformed, deviating from a homogeneous state, thus indicating the presence of non-uniform internal stresses. In fact, the observation that the free edges of the fabric spontaneously depart out of the $(x,y)$-plane by winding is a signature of an in-built prestresses. Moreover, for $L_w<L^0_w$ the inter-yarn contacts are not established everywhere, thus stitches can slide without further deformation, providing the fabric with very low stiffness. Thanks to the affine behavior, the prestressed state associated with the vector field $\vec{u}_0$ can be interpreted as the result of an elongation of the fabric from an absolute, homogeneous reference configuration for which $L_w\equiv L^*_{w}<L_w^0$ and $L_c\equiv L^*_{c}<L_c^0$ where all the stitches have the same size. This allows us to describe the position field of the stitches $\vec{u}$ as lateral and longitudinal displacements from this homogeneous state. Assuming that the size of a stitch is very small compared to the size of the fabric, the displacement field can be written as function of continuous space variables $(x,y)$:
\begin{equation}
\vec{u}(x,y)=\left|{\begin{array}{c} x+a_0(x,y)+\varepsilon a_1 (x,y)\\ \\ y+b_0(x,y)+\varepsilon b_1 (x,y)\\ \end{array}}\right.\;,
\label{eq:Posfield}
\end{equation}
where $(a_0,b_0)$ is an inhomogeneous displacement field that quantifies the deviation from the absolute reference configuration $|x|\leq \frac{L_c^*}{2}$ and $|y|\leq  \frac{L_w^*}{2}$. The affine trajectories of the stitches for $\varepsilon>0$ motivate the decomposition of the displacement field as an embedded deformation induced by the knitting process on top of a linear response to the applied strain. Moreover, the experimental results show that all the fields involved in Eq.~(\ref{eq:Posfield}) are slowly varying functions in space, so that the components of their spatial gradients are small compared to $1$.

\section{Stitch based model}
\subsection{Kinematics} 
\begin{figure}[tbh]
\centerline{\includegraphics[width=0.9\linewidth]{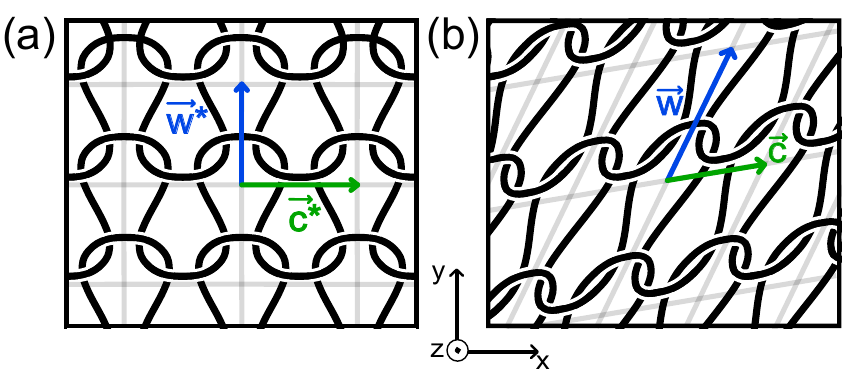}}
\caption{Schematic representation of the absolute reference configuration and a deformed state of the fabric. A geometrical representation of the stockinette stitch network consists in joining the geometric centers of neighbouring stitches along the wale and course directions. This defines a four-degree planar graph (light gray lines) that is uniquely determined by the edge vectors along the course and wale direction. (a) Absolute reference state of the fabric: the vectors $\protect\vec{c^*}$ and $\protect\vec{w^*}$ are orthogonal and identical for all cells. (b) Deformed state of the fabric, directions and norms of the vector fields $\protect\vec{c}$ and $\protect\vec{w}$ vary across the fabric.
\label{fig:geometry}}
\end{figure}

A knit is made of a single yarn that follows a complex constrained path. As the yarn runs along a given row, it alternatively intertwines with the top and bottom adjacent rows. To derive the morphological and mechanical properties of the whole fabric, we do not base our model on the yarn itself but rather on the periodic geometry of the knit. The fabric can be seen as a network of repetitive unit cells characterized by the yarn self-crossing topology. The stockinette pattern shows the advantage of having a single topology such that each stitch can be  associated to a unit cell  (see Fig.~\ref{fig:geometry}). We describe each cell by two vectors, $\vec{c}$ and $\vec{w}$, whose orientations prescribe the course and wale directions while their norms $c=\left\|\vec{c}\right\|$ and $w=\left\|\vec{w}\right\|$ impose the local dimensions of the cell. Notice that the yarn is not attached to these topological units and is allowed to slide from one stitch to another.

We can now write the constraints on the fabric that account for the permanence of the stitch topology. Indeed, whatever deformation we impose to the fabric, the crossing points cannot interchange. Therefore the stitch grid cannot lose or exchange cells and every stitch always keeps the same neighbours. This property imposes the following kinematic condition $\vec{w}(j,i)+\vec{c}(j,i+1)~=~\vec{w}(j+1,i)+\vec{c}(j,i)$. It simply states that, to go from one stitch to another, traveling along the rows and then the columns is equivalent to traveling along the columns and then the rows. Within a continuous representation, this constraint can be stated as: 
\begin{equation}
\frac{1}{w^*}\frac{\partial\vec{w}}{\partial x}(x,y)~=~\frac{1}{c^*}\frac{\partial\vec{c}}{\partial y}(x,y)\;,
\label{eq:constraint}
\end{equation} 
where $c^*$ and $w^*$ are the norms of the corresponding vectors in the absolute reference state of the fabric in which all the stitches have the same size (see Fig.~\ref{fig:geometry}(a)). Interestingly, Eq.~(\ref{eq:constraint}) can be seen as a component-wise vanishing divergence, which insures the existence of a vector potential $\vec{u}$ defined by:
\begin{equation}
\vec{c}=c^*\frac{\partial\vec{u}}{\partial x}\quad
\mbox{and}\quad\vec{w}=w^*\frac{\partial\vec{u}}{\partial y}\;.
\label{eq:def_cw}
\end{equation}
For a given $x$ (column) and $y$ (row), $\vec{u}(x,y)$ simply gives the position of the corresponding stitch, so this potential is identical to the position field introduced experimentally in Eq.~(\ref{eq:Posfield}). We can thus directly relate the vector fields $\protect\vec{c}$ and $\protect\vec{w}$ to the displacement fields $(a_0,b_0)$ and $(a_1,b_1)$.

\subsection{Energy} 
Our mechanical model proceeds from first-principle energy minimization under the constraint of fixed yarn length and a steady topology. Within the framework of our grid based model, we should express both elastic energy and the constraints as functions of $\vec{c}$ and $\vec{w}$. In the general case, the yarn would undergo stretching, bending and twisting. However, these deformation modes do not contribute equally to the energy of the fabric. First, the assumption of a scale separation between the size of a stitch and the diameter of the yarn, which corresponds to a rather \textit{loose} knit, allows us to invoke slender body approximation in which the energy cost of stretching is very large compared to that of bending. Moreover, the twist in the yarn does not change upon deformation of the stitch in the $(x,y)$ plane and thus will not produce any work. Therefore, the main contribution to elastic energy is provided by bending of the yarn and can be estimated using simple geometric arguments.

\begin{figure}[tbh]
\centerline{\includegraphics[width=0.9\linewidth]{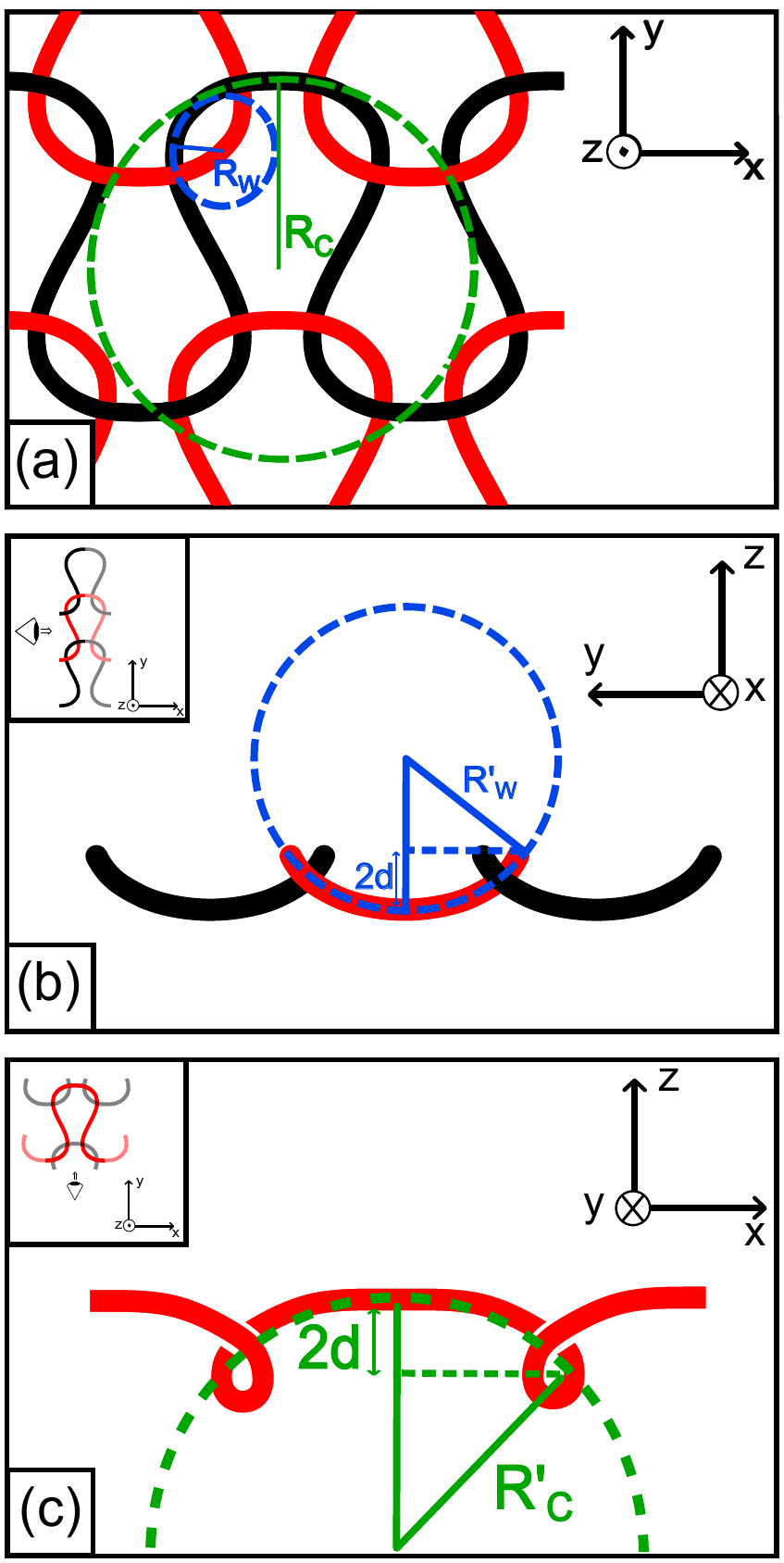}}
\caption{The main characteristic curvatures of the yarn that can be extracted from the geometry of a stockinette stitch. Panels (a) to (c) show the three different orthogonal planes.
\label{fig:energy}}
\end{figure}

In the $(x,y)$-plane of a given stitch, we can distinguish two characteristic radii of curvature $R_c$ and $R_w$ along the wale and course directions (see Fig.~\ref{fig:energy}(a)). Those curvatures are geometrically correlated to the dimensions $c$ and $w$ of the stitch. If we assume simple proportionality relations $R_c\sim c$ and $R_w\sim w$, the bending energies associated with deformations along the course (resp. wale) direction scale as $\mathcal{E}_c\sim B/c$ (resp. $\mathcal{E}_w\sim B/w$), where $B$ is the bending modulus of the yarn.  Orthogonally to $(x,y)$-plane, the yarn is also bent with two characteristic radii of curvature $R'_c$ and $R'_w$ (see Fig.~\ref{fig:energy}). The thickness of the fabric scales with the diameter of the yarn $d$, thus one has $R'_c\sim R_c^2/d$ and  $R'_w\sim R_w^2/d$. We consider here a slender yarn in a loose fabric, so that $R_c\ll R'_c$ and $R_w \ll R'_w$, with the result that bending energy carried in the orthogonal planes is negligible compared to that within the fabric plane. Nevertheless, $R'_c$ and $R'_w$ are responsible for the three dimensional shape that naturally occurs in an unloaded knitted fabric, in particular for the curling of the free edges of a stockinette sample. These out of plane effects do not directly impact the mechanical response of a clamped fabric but play a significant role in building up internal stresses. In our two-dimensional setting, the elastic energy of the stitch can thus be approximated by  
\begin{align}
\mathcal{E}_s 
&\approx \mathcal{E}_c+\mathcal{E}_w=\tilde B\left({\frac{1}{c}+\frac{\beta}{w}}\right)\;,
\label{eq:BendingEnergy}
\end{align} 
where $\tilde B$ is an effective bending modulus and $\beta$ is a dimensionless factor that quantifies the asymmetry between bending energies carried along the course and wale directions.

\subsection{Conservation of yarn length} 
Unlike a weaved fabric which can modelled by a Chebychev net for which all the edges retain fixed lengths~\cite{ghys2011coupe}, the deformation of a knitted fabric allows for sliding of the yarn from one stitch into the adjacent ones. Nevertheless, the assumption that the yarn experiences only bending deformations imposes that its total length in the fabric is conserved. The length of the yarn is correlated with the cumulated lengths of the edges of the associated network defined by the vectors $(\vec{c},\vec{w})$. In each cell, the yarn length should be proportional to $\ell=c+\delta w$, with $\delta$ a constant that embodies the complex geometry of the yarn in the stitch. 

\begin{figure}[tbh]
\centerline{\includegraphics[width=0.9\linewidth]{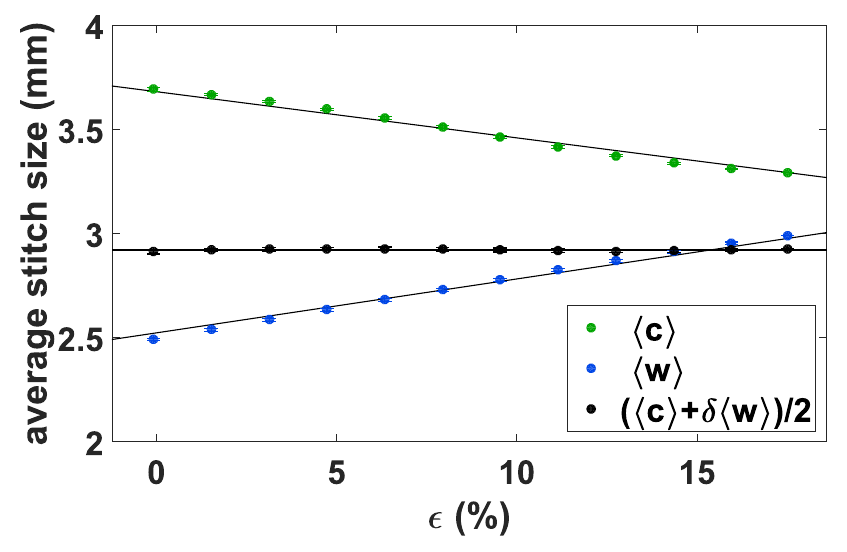}}
\caption{The average stitch size as function of the applied strain $\varepsilon>0$ of the fabric used in Fig.~\ref{fig:experiments}. It is shown that $\langle{c}\rangle$ (resp. $\langle{w}\rangle$) follows a negative (resp. positive) linear trend, thus we can define a parameter $\delta=0.86$ such that $\langle{c}\rangle+\delta\langle{w}\rangle$ is constant for all $\varepsilon>0$. Each data point is the average over 5 images at a given strain and the error bars are twice their standard deviation. 
\label{fig:measure_delta}}
\end{figure}

We assume beforehand that $\delta$ is a spatially uniform material parameter that is solely imposed by the machining process of the fabric. Therefore, the constraint on yarn length amounts to require that the average effective length over all stitches $\langle\ell\rangle =\langle c\rangle+\delta\langle w\rangle$ remains constant upon deformation. Using Eq.~(\ref{eq:def_cw}), the average sizes $\langle{c}\rangle$ and $\langle{w}\rangle$ can be determined from the experimentally measured deformation fields $\vec{u}(i,j)$. Fig.~\ref{fig:measure_delta} shows that both $\langle{c}\rangle$ and $\langle{w}\rangle$ are affine functions of $\varepsilon$ for all $\varepsilon>0$. Therefore, one can tune the parameter $\delta$ to prescribe a constant effective length $\langle\ell\rangle=\ell^*$. For our fabric, one finds $\delta=0.86$, which yields $\ell^*=5.86\,\mathrm{mm}$. This result justifies a posteriori our assumption that $\delta$ is a material parameter that characterizes the fabric geometry independently of the applied strain.

In the absolute reference state, the homogeneous rectangular configuration as defined by $c^*$ and $w^*$ should also satisfy the yarn length constraint, that is $c^*+\delta w^*=\ell^*$. A second equation relating $c^*$ to $w^*$ can be deduced by using the affine behavior of the vector field $\vec{u}$ as function of $\varepsilon$ (see Fig.~\ref{fig:stitch_track}) and assuming slowly varying deformation fields (see Appendix A). We find that the absolute reference state of the fabric used in Fig.~\ref{fig:experiments} is given by $c^*=3.93\,\textrm{mm}$ and  $w^*=2.08\,\textrm{mm}$. It is noteworthy that the conservation of yarn length combined with the experimental characterization of the local deformation field of the stitches allows us to determine both the geometric parameter $\delta$ and the absolute reference state. Moreover, using these quantities we can estimate the coupling of deformations in orthogonal directions by defining a geometric Poisson's ratio $\nu\equiv\frac{c^*-\langle{c}\rangle}{c^*}\frac{w^*}{\langle{w}\rangle-w^*}=\delta\frac{w^*}{c^*}=0.46$. Quite unexpectedly, even though the knitted fabric is very hollow, it behaves like an incompressible elastic bulk material with a conserved effective area.

\section{Elastic response of the fabric}

\subsection{Homogeneous deformations}

Let's first investigate the case of a knit that is uniformly deformed from its absolute reference configuration. Stretching the fabric in the wale direction by an amount $\varepsilon_h$ results in a deformation of each stitch given by  $w=w^*(1+\varepsilon_h)$. Using the conservation of yarn length, the deformation in the course direction should read $c=c^*(1-\delta \frac{w^*}{c^*}\varepsilon_h)$. Therefore, the bending energy of the fabric is simply given by
\begin{align}
\mathcal{E}_{h} &= \tilde B N_c N_w \left({\frac{1}{c^*(1-\delta \frac{w^*}{c^*}\varepsilon_h)}+\frac{\beta}{w^*(1+\varepsilon_h)}}\right)\,,
\label{eq:en_hom}
\end{align}
where $N_c$ and $N_w$ are the number of stitches in the course and wale direction. Notice that the asymmetry of bending energies between course and wale directions results both from the fact that the reference state is rectangular ($c^*\neq w^*$) and that the two directions of the stitch contribute to the yarn length in different proportions of its size. Eq.~(\ref{eq:en_hom}) shows that the stiffening behavior observed in the experiment is directly recovered: the elastic energy diverges as $\varepsilon_h\rightarrow \left(c^*/\delta w^*\right)$. More importantly, this energy should reach equilibrium at the absolute reference state $\varepsilon_h=0$. This condition allows us to prescribe $\beta=\delta\left(w^*/c^*\right)^2\approx0.24$. Similarly to $\delta$, the asymmetry parameter $\beta$ should be spatially homogeneous and marginally dependent on the applied deformation. Thus, $\beta$ can also be considered as a characteristic of the stitch.

\begin{figure}[tbh]
\centerline{\includegraphics[width=0.9\linewidth]{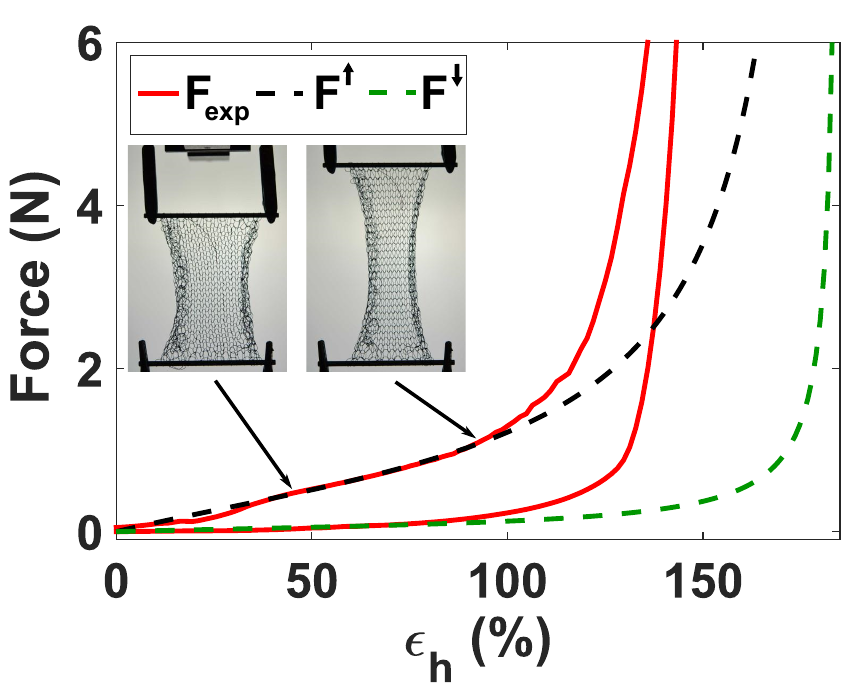}}
\caption{A $20\times 30$ stitches fabric is attached by its lower and upper rows to almost frictionless bars such that the stitches can slide laterally. The experimental pulling force $F_{exp}$ is averaged over 5 cycles. The absolute reference state $F_{exp}(\varepsilon_h =0)=0$ is determined by the position where loading and unloading curves coincide. The mechanical response $F(\varepsilon_h)$ is computed using Eq.~(\ref{eq:en_hom}) with $\tilde Y$ as the only fitting parameter to the experimental results. Because of friction within the knitted fabric, loading and unloading phases have different effective stretching moduli: $\tilde{Y}^{\uparrow}=3.8\times 10^{-2}\, \textrm{J.m}^{-1}$ and $\tilde{Y}^{\downarrow}=0.8 \times10^{-2}\, \textrm{J.m}^{-1}$.
\label{fig:homogeneous_def}}
\end{figure}

To test the relevance of Eq.~(\ref{eq:en_hom}), we have performed tensile tests using a loading configuration in which the knitted fabric is submitted to a quasi-homogeneous deformation. To this purpose, a knit made of $N_c\times N_w=20\times 30$ stitches is held by almost frictionless bars at its lower and upper rows such that the stitches can slide laterally. Movie S3 in Supplemental Material and Inset of Fig.~\ref{fig:homogeneous_def} show that the resulting catenary shape of the fabric is much less pronounced than in the case of clamped boundary conditions, confirming that the fabric deforms almost homogeneously for a large range of applied strains. For this case, one can define the deformation of the fabric with respect to the absolute reference configuration. Moreover, using Eq.~(\ref{eq:en_hom}) one can explicitly derive the elastic response of the fabric 
\begin{align}
F(\varepsilon_h)=-\frac{\partial \mathcal{E}_{h}}{\partial L_w}= \tilde{Y} N_c \delta \frac{w^*}{c^*}\left[\frac{1}{(1+\varepsilon_h)^2}-\frac{1}{(1-\delta \frac{w^*}{c^*}\varepsilon_h)^2}\right]
\label{eq:force_hom}
\end{align}
where the identities $\beta=\delta\left(w^*/c^*\right)^2$ and $L_w=N_w w^* (1+\varepsilon_h)$ have been used. The parameter $\tilde Y=\tilde B/(c^*w^*)$ is an effective stretching modulus that should be determined experimentally from the mechanical response of the fabric. Notice that $\tilde Y$ is a line tension (of dimension $\textrm{J.m}^{-1}$), which emphasizes that the mechanical response of the fabric originates from the tensions exerted on the entangled yarn. The parameters $\delta$, $w^*$ and $c^*$ should be determined from the conservation of total yarn length. Using the procedure described in Sec.~IIC, we find $\delta=0.71$, $w^*=2.1\,\mathrm{mm}$ and $c^*=2.7\,\mathrm{mm}$. Fig.~\ref{fig:homogeneous_def} shows the experimental mechanical response compared to the one given by Eq.~(\ref{eq:force_hom}). Fits to the linear regimes ($\varepsilon_h\ll1$) in the loading ($\uparrow$) and unloading ($\downarrow$) phases yield $\tilde{Y}^{\uparrow}=3.8\times 10^{-2}\, \textrm{J.m}^{-1}$ and $\tilde{Y}^{\downarrow}=0.8 \times10^{-2}\, \textrm{J.m}^{-1}$.

Finally, Fig.~\ref{fig:homogeneous_def} also shows that in this favorable loading configuration, the model allows us to keep track of the mechanical response up to more than $100\%$ deformations.  However, the sharp rise of the predicted force occurs at deformations $\varepsilon_h\sim c^*/(\delta w^*)$ which is larger than the experimental ones. Obviously at this range of deformations, one cannot invoke scale separation between the size of a stitch and the diameter of the yarn, a necessary condition to approximate the elastic energy of the stitch by Eq.~(\ref{eq:BendingEnergy}).

\subsection{Inhomogeneous deformations}

Before moving on to the inhomogeneous spatial deformation problem, let us summarize the parameters introduced in the model. There are mainly four internal geometrical parameters: $w^*$, $c^*$, $\beta$ and $\delta$, and one parameter related to the mechanics, the effective stretching modulus $\tilde{Y}$. As far as geometry is concerned, $w^*$ and $c^*$ are the extensions of the stitch at rest, thus specify the length scales of the microstructure and can be viewed as inputs. $\beta$, which represents the curvature asymmetry of the loops in the stitch, is determined from the mechanical equilibrium of the stitch at rest. $\delta$, which accounts for the asymmetry of thread length contributions to the stitch extensions in the course and wale directions, is recovered from thread length conservation. So at this point, there are no adjustable parameters other than a global stiffness scale $\tilde{Y}$.

Now, we have all the ingredients to study the mechanical response of the knit for any loading conditions. Within the experimental configuration of Fig.~\ref{fig:experiments} and using a continuous representation, the Lagrangian $\mathcal{L}\{\vec{c},\vec{w}\}$ of the fabric reads
\begin{widetext}
\begin{align}
\mathcal{L}&=\tilde Y{\iint\limits_{x,y}\mathrm{d}x\mathrm{d}y\left({\frac{1}{c}+\frac{\beta}{w}}\right)+\alpha \iint\limits_{x,y}\mathrm{d}x\mathrm{d}y\left({c+\delta w}\right) }-\iint\limits_{x,y}\mathrm{d}x\mathrm{d}y\,\vec{T}(x,y).\left({\frac{1}{c^*}\frac{\partial\vec{c}}{\partial y}-\frac{1}{w^*}\frac{\partial\vec{w}}{\partial x}}\right)-\iint\limits_{x,y}\mathrm{d}x\mathrm{d}y\, {\cal T}(x)\vec{e}_y.\vec{w}\;.
\label{eq:energy_tot}
\end{align}
\end{widetext}
The first term in Eq.~(\ref{eq:energy_tot}) is the elastic energy of the whole fabric, the second one ensures yarn length conservation, the third one enforces the local topological constraint and the last one is the work of the tractions exerted by external loads at the boundaries $y=\pm L^*_{w}/2$; $\alpha$, $\vec{T}(x,y)$ and ${\cal T}(x)$ being the corresponding Lagrange multipliers. One can interpret $\alpha$ as a global scaling for the tension in the fabric (see Appendix B), while $\vec{T}(x,y)$ is a local tension and ${\cal T}(x)$ is the applied traction along the clamped edges. 

\begin{figure}[tbh]
\centerline{\includegraphics[width=0.9\linewidth]{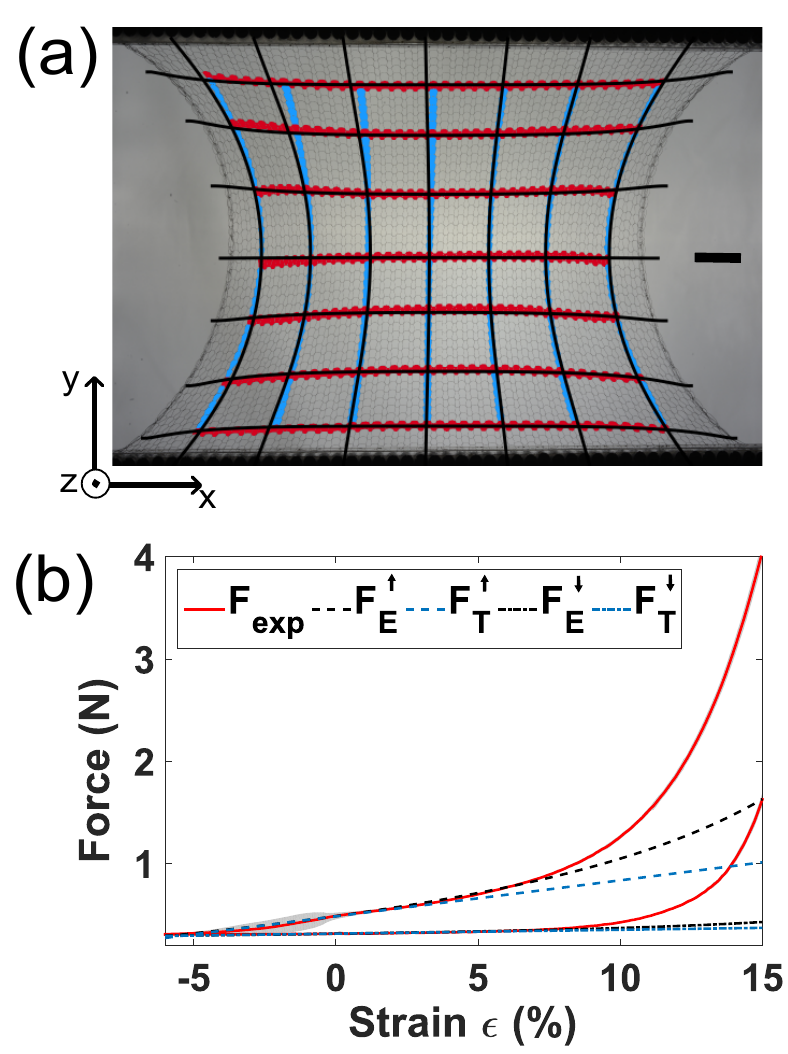}}
\caption{Predicted and experimental deformations of a stretched fabric in the $y$-direction and its mechanical response. (a) The measured displacement field $\vec{u}(x,y)$ is displayed along $7$ course (\textcolor[rgb]{0.8314,0,0.1137}{$\mathbf{-}$}) and $7$ wale (\textcolor[rgb]{0.0941,0.6039,0.9922}{$\mathbf{-}$}) directions, while the black curves ($\mathbf{-}$) are their corresponding predictions without adjustable parameters. The applied strain in the picture is $\varepsilon=11\%$ and the scale bar in the panels indicates $2\mathrm{cm}$. (b) Red curves reproduce the experimental results of Fig.~\ref{fig:experiments}(b). The grey area represents 2 times the standard deviation of the force signals over the 5 cycles. The forces $F_E$ (\textbf{-\,-}) and $F_T$ (\textcolor{bluematlab}{\textbf{$\cdot -$}}) are calculated respectively from the variation of bending energy and the Lagrange multiplier ${\cal T}(x)$ (see Appendix~C).
\label{fig:mech}}
\end{figure}

In Appendix B, we show that the minimisation of the Lagrangian with respect to the two vector fields $\vec{c}$ and $\vec{w}$ combined with the local topological constraint yields two Euler-Lagrange equations which can be expressed in terms of the displacement fields $a_0,b_0,a_1$ and $b_1$. Since these fields are slowly varying functions in space, only terms that are linear in their first spatial derivatives are retained. Then, the problem is solved in the limit of small strains and the corresponding displacement fields are computed. Fig.~\ref{fig:mech}(a) shows the resulting shape of the fabric compared to the experimental one. It is worth underlining that the prediction of the morphology of the fabric is parameter-free once the geometrical material parameters $\delta$, $\beta$, $c^*$, $w^*$ and the Lagrange multiplier $\alpha$ are determined.  Moreover, although we assume a small strain approximation, a fairly good quantitative agreement is found  between the results of our model and experiments for stretching up to $\varepsilon=15\%$.

Once the morphology of the fabric is computed, the force applied at $y=\pm L^*_{w}/2$ can be determined up to a scaling constant, using either the Lagrange multipliers $\vec T(x,y)$ and ${\cal T}(x)$, or the elastic energy of the fabric (Appendix C). The former ($F_T$) yields the affine behavior of the force at small strains and the latter ($F_E$) includes nonlinearities in the strain as the elastic energy is a nonlinear function of $c$ and $w$. The comparison with the measured force at small strains allows for the estimation of the effective stretching modulus $\tilde Y$. For the fabric used in Fig.~\ref{fig:experiments}, we find $Y^{\uparrow} \approx 1.5 \times 10^{-2}\textrm{J.m}^{-1}\approx 10^{-2} B/d^2$ for loading and $Y^{\downarrow} \approx 1.5 \times 10^{-3}\textrm{J.m}^{-1}\approx  10^{-3} B/d^2$ for unloading phases ($B$ is the bending modulus of the yarn and $d$ its diameter), showing that the fabric is very stretchable compared to its constituent yarn. Fig.~\ref{fig:mech}(b) shows that though linearized, our model allows for a reasonable prediction of the mechanical response up to $5\%$ deformation. Notice that for a clamped fabric, the applied strain $\varepsilon$ is defined with respect to an intermediate prestressed state described by the deformation field $\vec{u}_0$, in contrast to the homogeneous case where the absolute configuration is accessible. This explains the strain scale difference between Fig.~\ref{fig:homogeneous_def} and \ref{fig:mech}(b). As in the homogeneous case, the rise of the predicted force at deformations much larger than that observed experimentally confirms that a jamming phenomenon occurs before the non-linear behavior of our model, which confirms the relevance of our linearized approach for more complex loading configurations.

Finally, the framework we developed can be used for any in-plane deformation of the fabric. In order to further assess the predictive power of our model, we have investigated two other different loading configurations: the first one consists in tilting the initial knit with respect to the axis of applied loading by an angle $\theta=25^{\circ}$ and the second one ascribes an initial shear to the knit before uniaxial loading. Snapshots of representative deformed configurations for the two tests are shown in Fig.~\ref{fig:shape}(a) and (b) and the corresponding equations for the model are featured in Appendix~D. Despite the symmetry breaking, our model performs with an accuracy comparable to that obtained in the symmetrical loading case.

\begin{figure}[tbh]
\centerline{\includegraphics[width=0.9\linewidth]{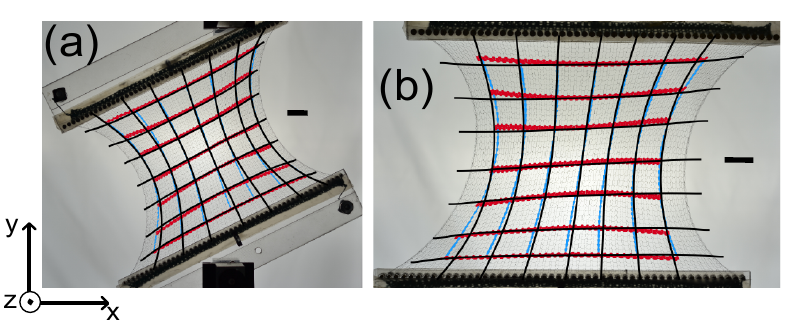}}
\caption{Predicted and experimental deformations of a stretched fabric under two different loading configurations. In (a) the fabric is initially tilted by $25^{\circ}$ and in (b) the fabric is initially sheared along the $x$-direction by an angle of $10^{\circ}$. The parameters $c^*$, $w^*$, $\delta$ and thus $\beta$ are the same than for the fabric used in Fig.~\ref{fig:mech}(a). The boundary conditions of these configurations are shown in Appendix~D. The applied strain is $\varepsilon=11\%$ for both fabrics and scale bars in the panels are $2\mathrm{cm}$.
\label{fig:shape}}
\end{figure}

\section{Discussion}
The experimental study of the mechanical response of a stretched fabric knitted into a stockinette stitch pattern allows us to build a reliable elastic model that recovers accurately its deformation field. The model  assesses the elastic energy of the fabric from the bending energy of its constituent yarn. Furthermore, the yarn self-crossing topology is represented as a $4$-degree planar graph that fully describes the state of the fabric. The topology imposes kinematic conditions, and yarn length conservation prescribes geometric properties of the resulting network. The displacement field of the stitches is then determined by constrained energy minimisation. The model correctly accounts for the spatial deformation of the fabric over a reasonable range of applied stretching and consistently captures the mechanical response upon setting a single material modulus $\tilde Y$ that differs depending on whether the fabric is stretched or relaxed. The results show that a knit behaves similarly to a rubberlike material: it is very stretchable and exhibits a geometric Poisson's ratio close to 0.5. Importantly, this analogy holds even though the material points of the underlying network are purely topological entities that do not correspond to bulk material points.

The equilibrium state for a stockinette pattern is not a flat surface, but rather a three dimensional configuration in which the fabric wraps around its edges due to out of plane bending of the yarn. The forced flattening and clamping of the fabric induce residual stresses and thus impose an inhomogeneous two-dimensional displacement field that deviates from an absolute homogeneous state. Such prestress would probably disappear from a fabric made of seed pattern, characterised by fully alternated knit and purl stitches~\cite{Anbumani2007knitting}, for which the equilibrium state is a flat surface. Nevertheless, our model is able to capture the equivalent two-dimensional residual deformation field of any fabric made of a periodic stitch pattern. This is in contrast with prestressed bulk elastic material for which the residual strain should be a prescribed field~\cite{klein2007shaping}.

Let us recall the role of friction in the morphology and mechanical response of the fabric. Friction is responsible for the large hysteresis observed in the dynamometry: in contrast to the elastic part of the response, friction opposes the deformation. This effect yields different stiffening behavior upon loading and unloading the fabric while keeping its morphology globally unaltered. Consequently, our measurements do not access the ``frictionless'' elastic modulus of the fabric but allows for the definition of two effective stretching moduli $\tilde Y^{\uparrow}$ for loading and $\tilde Y^{\downarrow}$ for unloading phase. It turns out that friction in a knitted fabric proceeds through stick-slip events and that the strong elastic recall forces bring the system back to its minimum energy configuration in high frequency, collective relaxation events of small amplitude. A rich phenomenology emerges from this frictional dynamics and the spatially extended avalanche-like relaxation events~\cite{Poincloux2018}.

Our model assumes scale separation between the yarn diameter and the stitch extension and thus applies as long as the yarn diameter is very small compared to stitches dimensions $c$ and $w$. Very often in commercial knits, the fluffiness of the thread gives the impression of tight stitches, but the fabric remains very stretchable. In these cases, such as for a scarf or a wool sweater, we expect the model to hold. Of course there are occurrences of tightly knitted fabric that do not feel any different from woven textile from a mechanical aspect; in such case our model would not be appropriate since deformation might involve significant thread elongation.
In addition, the so-called jamming of the stitches~\cite{hepworth1976mechanics} could occur when this scale separation no longer holds because of large local deformations. In those stitches, the yarn is maximally bent and starts to undergo stretching which alters the elasticity of the whole fabric. Even though jamming is localised in few stitches, the mechanical response of the knit becomes dominated by stretching of the yarn. Since the deformation field is inhomogeneous in the fabric, this phenomenon could occur locally even for small strains starting mainly from the corners. This mechanism is responsible for the nonlinear behavior of the force-strain curve (see Fig.~\ref{fig:mech}(b)) that is not captured by our model. However, it does not seem to affect the overall catenary shape of the fabric as drastically, which is still quantitatively predicted by the model. Indeed, we observe that the prediction of the model for the shape holds for lager deformation than that for the mechanical response. 

\section{Materials and Methods} 
The fabrics were crafted using a Toyota KS858 single bed knitting machine. All samples used for inhomogeneous deformations experiments were composed of $51\times51$ stockinette stitches made of a nylon-based monofilament (Stroft$^{\mbox{\scriptsize{\textregistered}}}$ GTM) of diameter $d=80\,\mu\textrm{m}$ and length of approximately $25\,\textrm{m}$. The yarn Young's modulus $E\approx 5.1\,\textrm{GPa}$ was measured using a tensile test, yielding a bending modulus $B\approx 10^{-8}\, \textrm{J.m}$. The fabrics were clamped at both extremities along the course direction by means of screws holding individually each stitch, imposing along the corresponding rows a constant spacing between the stitches. For homogeneous deformation experiments, a fabric of $20\times30$ stockinette stitches made of the same nylon-based yarn but of diameter $d=200\,\mu\textrm{m}$ was attached by its lower and upper rows to cylindrical steel bars such that the stitches can slide laterally. The steel bars were lubricated with silicon oil to reduce friction with the knit as much as possible. In all experiments, fabrics were stretched in the wale direction (except for the configuration of Fig.~\ref{fig:shape}(b)) at a constant speed of $0.1\,\textrm{mm/s}$ using an Instron$^{\mbox{\scriptsize{\textregistered}}}$ (model 5965) mounted with a $50\,\textrm{N}$ load cell. Starting from an initial configuration with a given $L^0_c$ and $L_{w}$, the fabrics were pulled on to a maximum distance of $30\,\textrm{mm}$. The visualisation was made using a Nikon$^{\mbox{\scriptsize{\textregistered}}}$ D800 camera with a $60\,\textrm{mm}$ 1:2:8:G AFS MicroNikkor lens. Both image and further data analysis were made using \textsc{Matlab} R2014b.  

\begin{acknowledgments}
We thank Cl\'ement Assoun for his precious advices in manufacturing the knitted samples, Sylvie M\'egret and Jean-Fran\c{c}ois Bassereau from the \'Ecole Nationale Sup\'erieure des Arts D\'ecoratifs for making available their knitting workshop and S\'ebastien Moulinet for fruitful discussions. This work was carried out in the framework of the METAMAT project  ANR-14-CE07-0031 funded by Agence Nationale pour la Recherche
\end{acknowledgments}


%

\begin{widetext}

\begin{appendix}

\section{Determination of the geometrical parameters $\delta$, $c^*$ and $w^*$}

The geometrical parameters of the knit can be directly measured from the position field of the stitches $\vec{u}(j,i)$. We can estimate the average value of the stitches size $\langle{c}\rangle$ and $\langle{w}\rangle$, with $\langle\rangle$ the average over all the stitches of the fabric for a given elongation (see Fig.~\ref{fig:measure_delta}).
It comes out from the measurements that for $\varepsilon>0$, both $\langle{c}\rangle$ and $\langle{w}\rangle$ vary linearly with $\varepsilon$. Thus we can define a simple scalar $\delta=0.86$ which generates $\ell ^*=\langle{c}\rangle+\delta\langle{w}\rangle=5.86\,\mathrm{mm}$ that is invariant with elongation. We then assume that $\ell ^*$ is proportional to the physical yarn length in the fabric. 

The linear trajectories of the stitches allows us to approximate the position field by $\vec u(j,i,\varepsilon) = \vec{u}_0(j,i)+\varepsilon\vec{u}_1(j,i)$. Moreover, we assume the existence of a reference configuration of the fabric where all the stitches have the same size $\vec c = c^* \vec{e}_x$  and $\vec w = w^* \vec{e}_y$. The reference configuration must also comply with the yarn length condition such that $\langle{c}\rangle+\delta\langle{w}\rangle=c^*+\delta w^*$. This equation in the limit of small deformation and slowly varying fields reads $\ell_0+\varepsilon \ell_1=c^* + \delta w^*$ with:
\begin{align}
\ell_0&=\frac{1}{N_c N_w}\iint\limits_{x,y}\frac{\mathrm{d}x\mathrm{d}y}{c^*w^*}(c^*(1+{a_0}_{,x})+\delta w^*(1+{b_0}_{,y}))=c^*+\delta w^*\;,
\label{eq:Cond0} \\
\ell_1&=\frac{1}{N_c N_w}\iint\limits_{x,y}\frac{\mathrm{d}x\mathrm{d}y}{c^*w^*}(c^*{a_1}_{,x}+\delta w^*{b_1}_{,y}+c^*{b_1}_{,x}{b_0}_{,x}+\delta w^*{a_1}_{,y}{a_0}_{,y})=0\;,
\label{eq:Cond1} 
\end{align}
where $N_c$ and $N_w$ are the number of stitches in the course and wale directions respectively and $f_{,x}=\frac{ \partial f}{\partial x}$. The size of the reference fabric is unknown and for this reason the experimental displacement fields $a_0$ and $b_0$ are functions of $c^*$ and $w^*$. 
The two previous equations can then be written as function of the experimental position fields $\vec u _0(i,j)$, $\vec u _1(i,j)$ and $c^*$,$w^*$. 
\begin{align}
&\ell_0=c^*+\delta w^* \iff \frac{1}{N_c N_w}\sum_{i,j}\left[{u_0^x(j+1,i)-u_0^x(j,i)+\delta\left({u_0^y(j,i+1)-u_0^y(j,i)}\right)}\right]=c^*+\delta w^* \;,
\label{eq:cond_length0}\\
&\ell_1=0 \iff\frac{1}{N_c N_w}\sum_{i,j}\left[{u_1^x(j+1,i)-u_1^x(j,i)+\delta\left({u_1^y(j,i+1)-u_1^y(j,i)}\right)}\right] = \label{eq:cond_length1}
\\ \nonumber
&-\sum_{i,j}\left[{(u_1^y(j+1,i)-u_1^y(j,i))\frac{u_0^y(j+1,i)-u_0^y(j,i)}{c^*}}{+\delta(u_1^x(j,i+1)-u_1^x(j,i))\frac{u_0^x(j,i+1)-u_0^x(j,i)}{w^*}}\right]\;,
\end{align}
with $\vec u_0=u_0^x\vec e_x+u_0^y\vec e_y$ and $\vec u_1=u_1^x\vec e_x+u_1^y\vec e_y$. Those equations can be solved to compute $c^*$ and $w^*$. For the the fabric used in Fig.~\ref{fig:experiments}, we find $w^*=2.08\,\textrm{mm}$ and $c^*=3.93\,\textrm{mm}$. The quantity $c^*+\delta w^*$ can then be compared to $\ell^*$ to verify the validity of our approximations. We measure $c^*+\delta w^*=5.70\,\textrm{mm}$ which generates less than $3\%$ of discrepancy, confirming the hypothesis of a reference knit sharing the same properties as the experimental knit and validating the approximation of small deformation and slowly varying displacement fields.

\section{Derivation of the equilibrium equations and solutions for the displacement fields}

When the deformation in the fabric is heterogeneous, solving the Lagrangian of the system Eq.~(\ref{eq:energy_tot}) provides with differential equations whose solutions ensure the energy minimization and the compliance to the constraints. Derivation of the Lagrangian with respect to $\vec c$ : $\mathcal{L}\{\vec c+\vec{\delta c},\vec w\}- \mathcal{L}\{\vec c,\vec w\}=0$, gives the following equation and its corresponding boundary condition.
\begin{align}
-\tilde Y\frac{\vec{c}}{c^3}+\alpha\frac{\vec{c}}{c}+\frac{1}{c^*}\frac{\partial\vec{T}}{\partial y}&=\vec 0 \label{eq:C}\\
\left[{\vec{T}.\vec{\delta C}}\right]_{y=\pm\frac{L^*_{w}}{2}}&=0
\label{eq:diff_c}
\end{align}
and with respect to $\vec w$ : $\mathcal{L}\{\vec c,\vec w+\vec{\delta w}\}- \mathcal{L}\{\vec c,\vec w\}=0$
\begin{align}
-\tilde Y \beta\frac{\vec{w}}{w^3}+\delta\alpha\frac{\vec{w}}{w}-\frac{1}{w^*}\frac{\partial\vec{T}}{\partial x}&={\cal T }(x) \vec {e}_y \label{eq:W} \\
\left[{\vec{T}.\vec{\delta W}}\right]_{x=\pm\frac{L^*_c}{2}}&=0
\label{eq:diff_w}
\end{align}
Notice that the dependance of $\alpha$ in Equations (\ref{eq:C},\ref{eq:W}) can be suppressed by normalizing $(\vec c,\vec w)$ by $\sqrt{\alpha}$ and $(\vec{T},\cal{T})$ by $\alpha$. This scaling allows us to interpret the Lagrange parameter $\alpha$ as a global scaling for the tension in the fabric. Moreover, Equations (\ref{eq:C},\ref{eq:W}) can be combined to eliminate the Lagrange multiplier $\vec T$:
\begin{equation}
c^*\frac{\partial}{\partial x}\left[{-\tilde Y \frac{\vec{c}}{c^3}+\alpha\frac{\vec{c}}{c}}\right] +w^*\frac{\partial}{\partial y}\left[{-\tilde Y \beta\frac{\vec{w}}{w^3}+\delta\alpha\frac{\vec{w}}{w}}\right]=0
\label{eq:MasterEquation}
\end{equation}
Using Equations~(\ref{eq:diff_c},\ref{eq:diff_w}), the boundary conditions corresponding to the experimental set-up of Fig.~\ref{fig:experiments} are $ \vec{c}(x,y=\pm\frac{L^*_{w}}{2})=\frac{L^0_c}{N_c}\vec{e}_x$ and $\int\frac{\mathrm{d}y}{w^*}\vec{w}=L_w\vec{e}_y$. The conditions on the free edges at $x=\pm\frac{L^*_c}{2}$ are $\vec T(x=\pm\frac{L^*_c}{2},y)=0$ that we can rewrite as a function of $\vec c$ thanks to Eq.~(\ref{eq:C}), $\left.{-\frac{\vec{c}}{c^3}+\alpha\frac{\vec{c}}{c} }\right|_{x=\pm\frac{L^*_c}{2}}=0$.
To guarantee the topological constraint, we write this equation and the boundary conditions as function of the displacements fields  $a_0$,  $b_0$, $a_1$ and $b_1$ thanks to the relations:
\begin{align}
\vec{c}&=c^*\left|{\begin{array}{c} 1+{a_0}_{,x}+\varepsilon{a_1}_x\\ {b_0}_{,x}+\varepsilon {b_1}_{,x}\\ \end{array}}\right. \\
\vec{w}&=w^*\left|{\begin{array}{c} {a_0}_{,y}+\varepsilon{a_1}_{,y}\\ 1+{b_0}_{,y}+\varepsilon{b_1}_{,y}\\ \end{array}}\right.
\end{align}
With the under-script $,x$ or $,y$ designating a partial derivative in the corresponding direction.
Experimental observations suggest those displacement fields being linear in $\varepsilon$, so that we also express the Lagrange multiplier $\alpha=\alpha_0+\varepsilon \alpha_1$ as varying linearly with $\varepsilon$.  Eq.~(\ref{eq:MasterEquation}) will be solved in the limit of small deformation and consequently developed at the first order of $\varepsilon$. We also consider that inner stress induces only a small deformation of the stitches, meaning that $a_{0_x},b_{0_x},a_{0_y},b_{0_y}\ll1$, thus the equation will also develop up to the first order in $a_{0_x},b_{0_x},a_{0_y},b_{0_y}$. By projecting Eq.~(\ref{eq:MasterEquation}) along $x$ and $y$ we end up with two sets of equations, each one with one part independent of $\varepsilon$ and another one proportional to the strain. Equilibrium of the homogeneous fabric for $c=c^*$ and $w=w^*$ imposes $\beta=\delta\left({\frac{w^*}{c^*}}\right)^2$. We define the dimensionless Lagrange multiplier $\tilde \alpha = \frac{{c^*}^2 \alpha}{\tilde Y}$ and introduce the two following coefficients to lighten the mathematical expressions, $\nu=\delta \frac{w^*}{c^*}$ and $\chi=\frac{\tilde{\alpha}_0-1}{2}$. For the independent part of the strain, one gets:
\begin{align}
{a_0}_{,xx}+\nu \chi{a_0}_{,yy}=&0 \\
{b_0}_{,xx}+\frac{\nu}{\chi}{b_0}_{,yy}=&0
\end{align}
Canceling the part proportional to $\varepsilon$ leads to:
\begin{align}
{a_1}_{,xx}+\nu \chi {a_1}_{,yy}+P_a=&0 
\label{eq:equation_affine1}\\
{b_1}_{,xx}+\frac{\nu}{\chi}{b_1}_{,yy}+P_b=&0
\label{eq:equation_affine2}
\end{align}
 With $P_a$ and $P_b$ terms that arise from the non-uniform initial state for $\varepsilon=0$, whose expressions are :
\begin{align}
P_a(x,y)=&-3\frac{\partial}{\partial x}\left[{{a_1}_{,x}{a_0}_{,x}}\right]+\nu\left({(1-\chi)\frac{\partial}{\partial y}\left[{{b_1}_{,y}{a_0}_{,y}+{a_1}_{,y}{b_0}_{,y}}\right]+\frac{\tilde\alpha_1}{2} {a_0}_{,yy}}\right)
\label{eq:perturbation_affine1} \\
P_b(x,y)=&(\frac{1}{\chi}-1)\frac{\partial}{\partial x}\left[{{b_1}_{,x}{a_0}_{,x}}\right] +\nu\left({\frac{\partial}{\partial y}\left[{(\frac{1}{\chi}-1){a_1}_{,y}{a_0}_{,y}-\frac{3}{\chi}{b_1}_{,y}{b_0}_{,y}}\right]}\right)
\label{eq:perturbation_affine2}
\end{align}
The boundary conditions written with the displacement fields are:
\begin{align}
\left.{\begin{array}{c}\forall x \\ y=\pm\frac{L^*_{w}}{2}\end{array}}\right.  \left\{{\begin{array}{l} a_0=\frac{L^0_c-L^*_c}{L^*_c}x\\\\ b_0=\pm\frac{L^0_w-L^*_{w}}{2} \\ \end{array}}\right.\,
\left.{\begin{array}{c}\forall y \\ x=\pm\frac{L^*_c}{2}\end{array}}\right. \left\{{\begin{array}{l} {a_0}_{,x}=-\chi \\\\ {b_0}_{,x}=0 \\ \end{array}}\right.
\end{align}
and
\begin{align}
\left.{\begin{array}{c}\forall x \\ y=\pm\frac{L^*_{w}}{2}\end{array}}\right. \left\{{\begin{array}{l} a_1=0\\\\  b_1=\pm\frac{L^0_w}{2} \\ \end{array}}\right. \quad
\left.{\begin{array}{c}\forall y \\ x=\pm\frac{L^*_c}{2}\end{array}}\right.\left\{{\begin{array}{l} {a_1}_{,x}=-\frac{\tilde\alpha_1}{2(3\chi+1)} \\ \\ {b_1}_{,x}=0 \\ \end{array}}\right.
\label{eq:bc_perturbation}
\end{align}

Let us start by the determination of $a_0,b_0$, the displacement from the absolute reference configuration.
The solution for $b_0$ is straightforward, it is simply a linear solution in $y$:
\begin{equation}
b_0(x,y)=\frac{L^0_w-L^*_{w}}{L^*_{w}}y
\end{equation}
The solution for $a_0(x,y)$ is more advanced, let us make the following change of variables
\begin{align}
x&=X\frac{L^*_c}{2}\\  
y&=Y \frac{\sqrt{\nu\chi}L^*_c}{2} \\
 a_0&=\frac{ L^*_c}{2}\left({\left({\chi+\frac{L^0_c-L^*_c}{L^*_c}}\right)A+\frac{L^0_c-L^*_c}{L^*_c}X}\right) \\ L&=\frac{L^*_{w}}{\sqrt{\nu\chi}L^*_c}
\end{align}
The equations for $a_0$ become
\begin{equation}
A_{XX}+A_{YY}=0
\end{equation}
with the boundary conditions
\begin{equation}
A(X,Y=\pm L)=0 \;\;\mbox{and}\;\; A_X(X=\pm 1,Y)=-1
\end{equation}
Therefore $A(x,y)$ is a harmonic function inside a rectangle that satisfies Dirichlet conditions at the boundaries. Semi-analytical resolution can be performed using conformal mapping techniques. In the present case, we numerically solve this problem using the Schwarz-Christoffel transform Matlab Toolbox~\cite{driscoll1996algorithm}.

Combining Eqs~(\ref{eq:equation_affine1},\ref{eq:equation_affine2}) and Eqs~(\ref{eq:perturbation_affine1},\ref{eq:perturbation_affine2}), one gets the following equations for the affine displacement field $(a_1,b_1)$ 
\begin{align}
& (1-3 {a_0}_{,x}){a_1}_{,xx}+\nu\chi\left({1+\left({\frac{1}{\chi}-1}\right){b_0}_{,y}}\right){a_1}_{,yy}+\nu(1-\chi){a_0}_{,y} {b_1}_{,yy} =
-\nu{a_0}_{,yy}\left({3\chi {a_1}_{,x}+(1-\chi){b_1}_{,y}+\frac{\tilde{\alpha_1}}{2}}\right) \\
& \left({1+\left({\frac{1}{\chi}-1}\right){b_0}_{,y}}\right){b_1}_{,xx}+\frac{\nu}{\chi}\left({1-3{b_0}_{,y}}\right){b_1}_{,yy}+\frac{\nu}{\chi}(1-\chi){a_0}_{,y} {a_1}_{,yy} =\frac{\nu}{\chi}(1-\chi){a_0}_{,yy}\left({\chi {b_1}_{,x}-{a_1}_{,y}}\right) 
\end{align}
These equations combined with the boundary conditions~(\ref{eq:bc_perturbation}) are solved numerically on a discrete lattice.

\begin{figure}[tbh]
\includegraphics[width=0.9\linewidth]{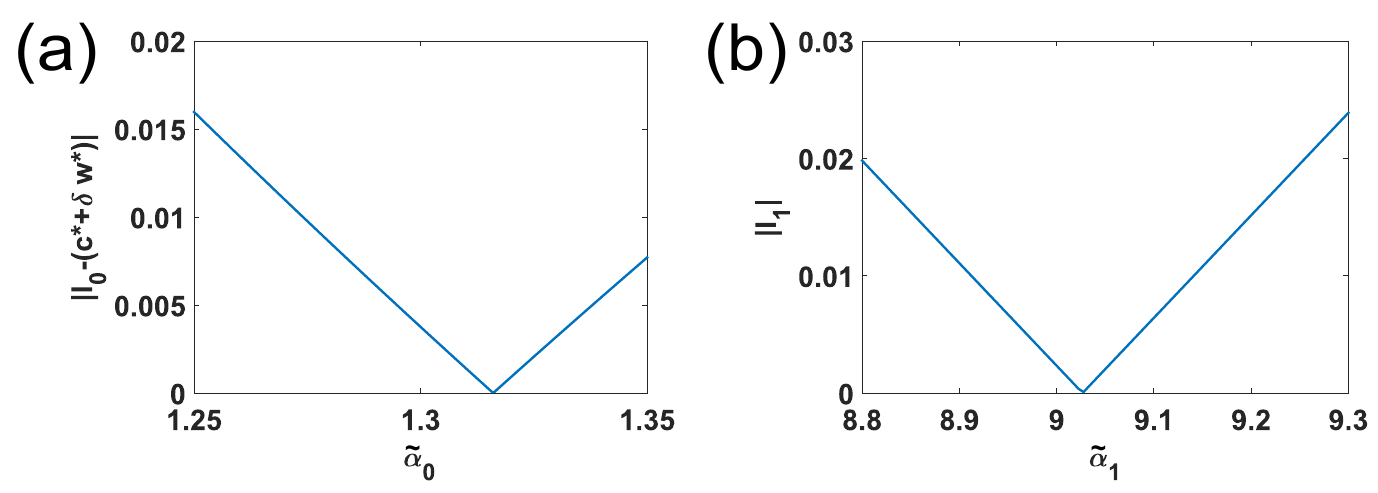}
\caption{Determination of the Lagrange parameter $\tilde{\alpha}=\tilde{\alpha}_0+\varepsilon\tilde{\alpha}_1$. (a) For each value of $\tilde{\alpha}_0$, $a_0$ and $b_0$ are calculated and $\ell_0$ is estimated, we find that $\ell_0-(c^*+\delta w^*)=0$ for $\tilde{\alpha}_0=1.32$. (b) For $\tilde{\alpha}_0=1.32$, while $\tilde{\alpha}_1$ is varied, $a_1$ and $b_1$ are calculated to estimate $\ell_1$, we find $\ell_1=0$ for $\tilde{\alpha}_1=9.04$.
\label{fig:condalpha}}
\end{figure}

Recall that the geometrical parameters of the knit $c^*$, $w^*$ and $\delta$ are computed from the measured position field of the stitches. However, the calculated displacement field is still parametrised by the Lagrange multipliers $\tilde{\alpha}_0$ and $\tilde{\alpha}_1$ which are the only parameters left to compute the shape of the fabric. $\tilde{\alpha}=\tilde{\alpha}_0+\varepsilon\tilde{\alpha}_1$ being the Lagrange multiplier associated to the yarn length conservation, we find its value using the corresponding equations $\ell_0=c^*+\delta w^*$ and $\ell_1=0$ but this time estimated from our model and not from experimental measurements, i.e one should satisfy Eqs~(\ref{eq:cond_length0},\ref{eq:cond_length1}) using the computed deformation vector field $\vec{u}(x,y)$. Notice that $\ell_0$ is independent of $\tilde{\alpha}_1$, thus we can compute $\tilde{\alpha}_0$ such that $\ell_0(\tilde{\alpha}_0)=c^*+\delta w^*$ (see Fig.~\ref{fig:condalpha}(a)). Then, with the selected $\tilde{\alpha}_0$, we apply the same treatment to $\ell_1$ and thus determine $\tilde{\alpha}_1$ that satisfies $\ell_1(\tilde{\alpha}_1)=0$ (see Fig.~\ref{fig:condalpha}(b)). For the fabric used in Fig.~\ref{fig:experiments}, we find $\tilde{\alpha}_0=1.32$ and $\tilde{\alpha}_1=9.04$.

\section{Computation of the Lagrange multiplier $\vec T(x,y)$ and the applied force}

We have introduced the vector $\vec{T}$ as the Lagrange multiplier associated with the topological constraint. Since we express the variables $\vec{c}$ and $\vec{w}$ directly in terms of displacement field, the topological constraint is automatically fulfilled and thus $\vec{T}$ does not appear explicitly in the solutions. However, once $a_0, a_1, b_0, b_1$ are determined, one can evaluate this vector field using Equations (\ref{eq:C},\ref{eq:W}), provided with the boundary condition $\vec{T}(x=\pm\frac{L^*_c}{2},y)=\vec{0}$. Recall that ${\cal T} (x) $ is a traction distribution such that the total force applied on the fabric writes
\begin{align}
F_T=w^*\int_{-\frac{L^*_c}{2}}^{\frac{L^*_c}{2}} {\cal T}(x)dx
\end{align}
Let us define a new vector $\vec{T}(x,y)=\tilde{Y}\vec{T}_1(x,y)$ and ${\cal T} (x)=\tilde{Y}{\cal T}_1 (x)/w^{*2}$. Equations (\ref{eq:C},\ref{eq:W}) become
\begin{align}
\frac{1}{c^*}\frac{\partial\vec{T}_1}{\partial y}&=\frac{\vec{c}}{c^3}-\frac{\tilde \alpha}{c^{*2}}\frac{\vec{c}}{c} \equiv \frac{1}{c^{*2}} \vec{f}_c(x,y)\label{eq:C1}\\
\frac{1}{w^*}\frac{\partial\vec{T}_1}{\partial x}&=-\delta \left(\frac{w^*}{c^*}\right)^2\left[\frac{\vec{w}}{w^3}-\frac{\tilde \alpha}{w^{*2}}\frac{\vec{w}}{w}\right] -\frac{1}{w^{*2}}{\cal T }_1(x) \vec {e}_y \equiv \frac{1}{w^{*2}} \vec{f}_w(x,y)-\frac{1}{w^{*2}}{\cal T }_1(x)\vec {e}_y\label{eq:W1} 
\end{align}
The determination of the the vector field $\vec{T}_1(x,y)$ allows us to compute the applied force. Integrating Eq.~(\ref{eq:W1}) and using the boundary condition $\vec{T}(x=-\frac{L^*_c}{2},y)=\vec{0}$ yields
\begin{align}
\vec{T}_1(x,y)= \frac{1}{w^*}\int_{-\frac{L^*_c}{2}}^x \vec{f}_w(x',y) \mathrm{d}x'-\frac{1}{w^*}\int_{-\frac{L^*_c}{2}}^x {\cal T }_1(x)\vec {e}_y \mathrm{d}x'
\end{align}
Using Eq.~(\ref{eq:MasterEquation}), one can show that this solution for $\vec{T}_1$ satisfies Eq.~(\ref{eq:C1}). Moreover, using the boundary condition $\vec{T}(x=\frac{L^*_c}{2},y)=\vec{0}$, one finds
\begin{align}
F_T= \tilde{Y}\int_{-\frac{L^*_c}{2}}^{\frac{L^*_c}{2}} \vec{f}_w(x,y)\cdot \vec {e}_y\frac{\mathrm{d}x}{w^*}
\end{align}
This shows the integral of the tension over the course direction should be a conserved quantity independent of the coordinate in the wale direction.

To remain consistent with the solutions obtained for $a_0,b_0,a_1$ and $b_1$, the vector field $\vec{f}_w(x,y)$ is developed to first order in $\varepsilon$ and in $a_{0_{,x}},b_{0_{,x}},a_{0_{,y}},b_{0_{,y}}$. To compute the tension field $\vec{T}$ and the force $F_T$ one needs to know the stretching modulus $\tilde Y$.
However, the force needed to pull the fabric can also be obtained from the variation of bending energy with respect to $L_w$This provides an alternative estimation of the force that we name $F_E$. Both $F_T$ and $F_E$ are proportional to the unknown effective stretching modulus $\tilde Y$. This coefficient is adjusted such that the slopes of $F_{exp}$ and $F_T$ near $\varepsilon=0$ coincide, we find $\tilde Y \approx 1.5 \times 10^{-2}\textrm{J.m}^{-1}$. The resulting three curves are shown in Fig.~\ref{fig:mech}(b). The consistency of the model and the correctness of the computations appear here as the slope of $F_E$ matches those of $F_{exp}$ and $F_T$ in the vicinity of $\varepsilon=0$. While $F_{exp}(\varepsilon)$ is well described by $F_T$ at small strains (up to $\varepsilon \approx 4\%$), the nonlinear behavior of $F_E$ manages to capture $F_{exp}$ until $\varepsilon \approx 8\%$. We apply the same process for the unloading phase and we find $\tilde Y \approx 1.5 \times 10^{-3}\textrm{J.m}^{-1}$.

The tension $\vec{T}(x,y)$ over the fabric cannot be retrieved independently of ${\cal T}(x)$. However, one can compute an equivalent total stress $\vec T _{tot}=\vec{T}+w^*\int\limits_0^x {\cal T}(x)\mathrm d x\,\vec{e}_y$, see Fig.~\ref{fig:T} for the amplitude of this vector field over the fabric. Notice that $\vec{T}$ is the Lagrange multiplier associated to the topological constraint, so one interpretation of $\Vert\vec{T}_{tot}\Vert$ would be how hard it is to fulfill this constraint. In other words, it can be interpreted as an inter-stitch tension and its behavior reflects a stress distribution within the fabric.

\begin{figure}[tbh]
\centerline{\includegraphics[width=0.45\linewidth]{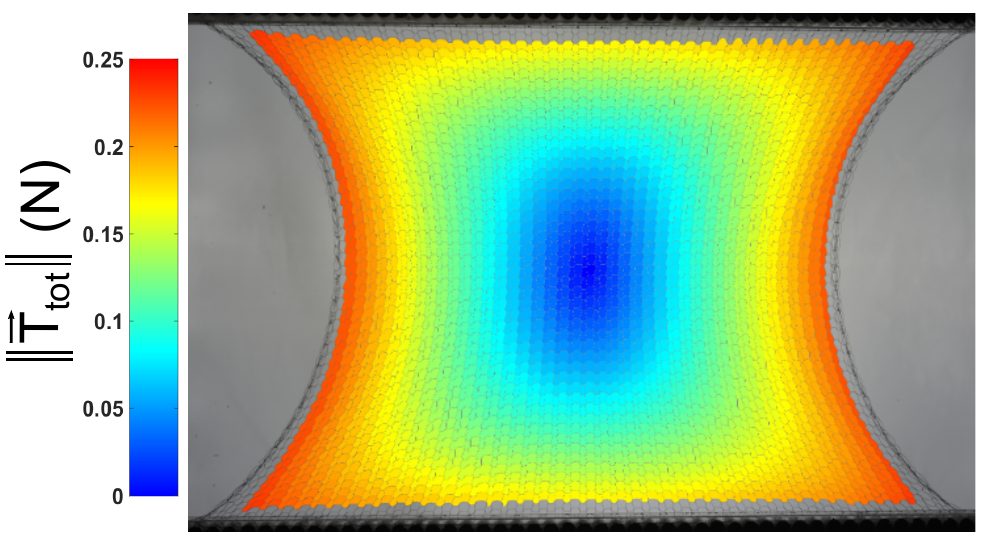}}
\caption{Amplitude of the vector Lagrange multiplier $\Vert\vec{T}_{tot}\Vert$ evaluated for $\varepsilon=11\%$ and displayed over the corresponding picture of the fabric, each stitch being colored by its value of $\Vert\vec{T}_{tot}\Vert$.}
\label{fig:T}
\end{figure}

\section{Equations for the alternatives loading conditions}

Fig.~\ref{fig:shape}(a) and (b) show the deformation fields for different loading configurations applied to the fabric. In the model, those changes will not affect the equilibrium equations of the displacement fields, but only the associated boundary conditions. In Fig.~\ref{fig:shape}(a) the fabric is tilted by an angle $\theta=25^{\circ}$ and loaded in the $y$-direction. The initial configuration is identical as the symmetrically loaded one, hence $a_0$ and $b_0$ satisfy the same equations with the same boundary conditions. For the affine displacement fields $a_1$ and $b_1$ the boundary conditions become:
\begin{align}
\left.{\begin{array}{c}\forall x \\ y=\pm\frac{L^*_{w}}{2}\end{array}}\right. \left\{{\begin{array}{l} a_1=\pm\frac{L^0_w}{2}\sin(\theta)\\\\  b_1=\pm\frac{L^0_w}{2}\cos(\theta) \\ \end{array}}\right. \quad
\left.{\begin{array}{c}\forall y \\ x=\pm\frac{L^*_c}{2}\end{array}}\right.\left\{{\begin{array}{l} {a_1}_{,x}=-\frac{\tilde\alpha_1}{2(3\chi+1)} \\ \\ {b_1}_{,x}=0 \\ \end{array}}\right.
\end{align}

For the second configuration displayed in Fig.~\ref{fig:shape}(b), the fabric is initially sheared by an angle of $\gamma=10^{\circ}$, thus inducing an additional lateral displacement. The boundary conditions associated to the prestressed state are then: 
\begin{align}
\left.{\begin{array}{c}\forall x \\ y=\pm\frac{L^*_{w}}{2}\end{array}}\right.  \left\{{\begin{array}{l} a_0=\frac{L^0_c-L^*_c}{L^*_c}x\pm \tan(\gamma) \frac{L^0_w}{2}\\\\ b_0=\pm\frac{L^0_w-L^*_{w}}{2} \\ \end{array}}\right.\,
\left.{\begin{array}{c}\forall y \\ x=\pm\frac{L^*_c}{2}\end{array}}\right. \left\{{\begin{array}{l} {a_0}_{,x}=-\chi \\\\ {b_0}_{,x}=0 \\ \end{array}}\right.
\end{align}
The equations and boundary conditions for $a_1$ and $b_1$ are functions of $a_0$ and $b_0$ so their solutions are different from the symmetric case, even if their expressions are identical.

\end{appendix}
\end{widetext}

\end{document}